\documentclass[twocolumn,aps,showpacs,preprintnumbers,amsmath,amssymb,superscriptaddress,nobibnotes]{revtex4}
\usepackage{graphicx}
\usepackage{dcolumn}
\usepackage{bm}
\usepackage{graphics}
\usepackage{longtable}
\usepackage{units}
\usepackage{textcomp}

\begin{document}

\title{Biprism Electron Interferometry with a Single Atom Tip Source}

\author{G. Sch\"{u}tz$^1$, A. Rembold$^1$, A. Pooch$^1$, S. Meier$^1$,
P. Schneeweiss$^2$, A.~Rauschenbeutel$^2$, A. G\"{u}nther$^3$, W.T. Chang$^4$, I.S. Hwang$^4$ and A. Stibor}

\affiliation{Institute of Physics and Center for Collective Quantum Phenomena in LISA$^+$,
University of T\"{u}bingen, Auf der Morgenstelle 15, 72076 T\"{u}bingen, Germany\\
$^2$Vienna Center for Quantum Science and Technology, TU Wien - Atominstitut, 1020 Vienna, Austria\\
$^3$Institute of Physics and Center for Collective Quantum Phenomena in LISA$^+$,
University of T\"{u}bingen, Auf der Morgenstelle 14, 72076 T\"{u}bingen, Germany\\
$^4$Institute of Physics, Academia Sinica, Nankang, Taipei, Taiwan, Republic of China}


\begin{abstract}
Experiments with electron or ion matter waves require a coherent, monochromatic and long-term stable  source with high brightness. These requirements are best fulfilled by single atom tip (SAT) field emitters. The performance of an iridium covered W(111) SAT is demonstrated and analyzed for electrons in a biprism interferometer. Furthermore we characterize the emission of the SAT in a separate field electron and field ion microscope and compare it with other emitter types. A new method is presented to fabricate the electrostatic charged biprism wire that separates and combines the matter wave. In contrast to other biprism interferometers the source and the biprism size are well defined within a few nanometers. The setup has direct applications in ion interferometry and Aharonov-Bohm physics. 
\end{abstract}

\maketitle

\section{Introduction}

In consideration of the success with interferometers for neutral atoms  \cite{Carnal1991a,Keith1991a,Gustavson1997a,Peters1999a} and molecules \cite{Grisenti2000a,Brezger2002a,Gerlich2007a}, efforts were made to expand the field towards matter wave experiments with ions. They started with the realization of the first biprism interferometer for helium ions by Hasselbach et al. \cite{Hasselbach1998a,Hasselbach1996,Maier1997,Hasselbach2010a}. The additional parameter charge combined with the internal structure of ions offer significant advantages in comparison to interferometers for electrons and neutral atoms. Ion interferometers could open up the door for a class of novel quantum optical experiments to test the influence of inner structure in the magnetic Aharonov-Bohm effect \cite{Aharonov1959,Silverman1993,Schutz2013} or the first direct measurement of the electric Aharonov-Bohm effect \cite{Schutz2013}. Furthermore such a device may allow for interferometry experiments with laser excitation of inner ionic states \cite{Silverman1993}, decoherence studies \cite{Sonnentag2007a} and might enable for extremely sensitive sensors for rotations and accelerations \cite{Clauser1988}.\\
However, in the first and only realization of an ion interferometer \cite{Hasselbach1998a,Hasselbach1996,Maier1997,Hasselbach2010a}, the low helium ion brightness of the field ion source turned out to be the weak point of the design. Very long exposure times of \unit[15]{min} had to be taken into account to accumulate interference fringes and diffraction for \unit[3]{keV} He$^+$ ions. For that reason further quantum optical experiments have not been possible.  A second attempt to build an ion interferometer was made by Krenn et al. \cite{Krenn1999}. Also in this device the low count rate of the ion source was the main challenge and prevented the successful generation of ion interferences. To solve this issue an alternative source needs to be applied. It was shown in \cite{Kuo2006a,Kuo2008,Fu2001,Chang2009,Hwang2013} that single atom tip (SAT) field emitters  generate coherent beams with high brightness for electrons and ions.\\
\\
In this article we demonstrate for the first time the application of an iridium covered W(111) SAT source in a biprism interferometer. The device is based on the setup of Hasselbach et al. \cite{Hasselbach1998a,Hasselbach1996,Maier1997,Hasselbach2010a}.
We analyze the electron interference pattern in this interferometer, characterize the performance of the SAT emitter in a field ion and field electron microscope and compare it with other emitter types. Furthermore, a new method to fabricate a coherent biprism beam splitter is presented. \\
In conventional biprism electron or ion interferometers \cite{Hasselbach1998a,Hasselbach1996,Maier1997,Hasselbach2010a,Sonnentag2007a,Mollenstedt1956a,Nicklaus1993a,Hasselbach1988a}  the source size and the biprism diameter are not known with high accuracy. In our setup these parameters are well defined within several nanometers. Recently an electron beam from a SAT source was diffracted and interfered on a carbon nanotube \cite{Chang2009,Hwang2013}. Although the high coherent properties of this source could be demonstrated, it is likely that this scheme cannot be applied for ion interferometry, since the fringe pattern magnification was performed geometrically with a distance of several hundred nanometers between the tip and the nanotube. For ions the tip voltage needs to be higher and the distance to the nanotube shorter, which may cause arcing. In this context our setup has the advantage that the interference pattern is magnified by a quadrupole lens. Furthermore the interference can be tuned by an adjustable potential on the biprism fiber, which is not possible in the nanotube approach.

\section{Biprism Interferometry}

In biprism interferometry \cite{Mollenstedt1956a} an ideally point-like electron or ion source illuminates coherently a biprism fiber. Applying a positive (for electrons) or negative (for ions) potential on the wire leads to a separation and deflection of the matter wave. The beam paths combine again shortly after the biprism and interfere with each other. As a result an interference pattern parallel to the biprism wire can be detected in the plane of observation.

\subsection{Single Atom Tip Field Emitters}\label{SAT}

The choice of the beam source with a highly coherent signal is crucial for an electron or ion biprism interferometer that should be capable of performing sophisticated future matter wave experiments. The lateral and longitudinal coherence length, the emission angle, the signal intensity and the spatial stability depend on the source. An established technique to produce a coherent electron beam is to etch a metal tip and set it on a high negative potential in vacuum until field emission starts. To generate an ion beam, a positive voltage is applied to ionize and accelerate gas atoms at the tip end due to the high electric field. \\
Several different tip types have been developed in the last decades. The electron emission characteristics of four of them are listed in Tab. \ref{tab:comparison}. In the first one, electrons are thermally emitted by a comparably blunt tungsten or lanthanum hexaboride (LaB$_6$) tip and accelerated towards a counter electrode \cite{Reimer1998}. The second type describes an etched tip where the electrons are extracted by field emission. The third one, the so called "supertip", is an etched tip, where a tiny protrusion is created \cite{Kalbitzer,Jousten1988}. This was the source in the first ion interferometer \cite{Hasselbach1998a,Hasselbach1996,Maier1997,Hasselbach2010a}. Due to its geometry, the emitted electron or ion matter wave experience a self focusing effect. An ion beam is emitted with a high angular confinement of about 2$^\circ$, instead of about 60$^\circ$ full emission angle for etched tips \cite{Knob} without the protrusion. However, it was mentioned in \cite{Maier1997} for ion emission that this protrusion, being the emission center, is spatially only stable for about one hour, with a large uncertainty between different supertips. This is a disadvantage for the use in ion interferometry since the beam alignment before signal acquisition is typically longer than this time and the instability leads to uncorrelated phase shifts, destroying the contrast. Also the maximal beam brightness of \unit[$10^{14}$]{A/m$^2$ sr} for ionization gas with a partial pressure of \unit[1]{mbar} (Tab. \ref{tab:comparison}, \cite{Boerret1990}) could not be reproduced in the ion interferometer of Hasselbach et al. \cite{Maier1997}. There the maximal helium beam brightness was in the range of \unit[$10^{12}$]{A/m$^2$ sr}.  This tip type is therefore not practical for further experiments with a larger beam separation and a long signal integration time, such as required for Aharonov-Bohm physics \cite{Aharonov1959}.

\begin{figure}[htbp]
\centerline{\scalebox{0.41}{\includegraphics{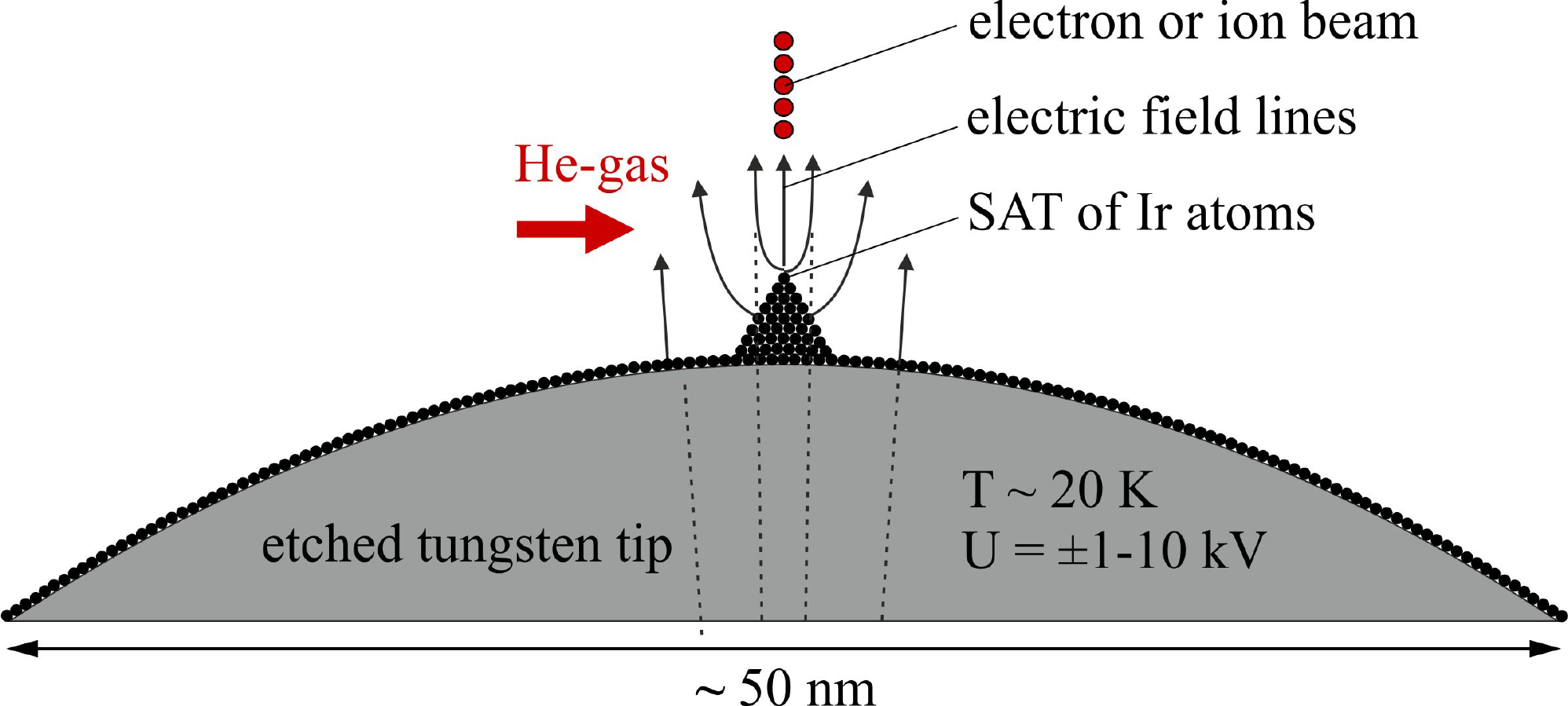}}}
\caption{(Color online) Sketch of field electron or field ion emission from a SAT as a coherent source for interferometry. The arrows indicate the direction of the electric field lines orthogonal to the tip surface resulting in a self focusing effect.} 
\label{fig:tip}
\end{figure}

\renewcommand{\arraystretch}{1.3}
\begin{table*}[htbp]
	\centering
		\begin{tabular}{|l||c|c|c|c|}	\hline
			& Thermal emission & Etched tip & Supertip & SAT	\\	\hline\hline
		Source diameter & \unit[20-100]{\textmu m} \cite{Reimer1998} & \unit[$<$10]{nm} \cite{Chang2012} & \unit[4]{nm} \cite{Kalbitzer} & $\unit[3]{\mathring{A}}$ \cite{Kuo2008}	\\	\hline
		Energy spread $[\unit[]{eV}]$ & $0.5-2$ \cite{Reimer1998} & 0.2 \cite{Reimer1998} & 0.5  \cite{Knob} (ions: 1 \cite{Borret88}) & 0.4 \cite{Kuo2006a} \\ \hline
		Max. brightness & $10^9$ \cite{Reimer1998} & $10^{13}$ \cite{Reimer1998} & $10^{13}-10^{14}$ \cite{Knob} & 	$10^{13}-10^{14}$ \cite{Ishikawa2007,Kuo2004}\\
		$[\unit[]{A/m^2 \, sr}]$ & && (ions: $10^{14}$ \cite{Boerret1990}) & (ions: $2 \times 10^{15}$ \cite{Kuo2008}) \\	\hline
		\end{tabular}
	\caption{Comparison of the source diameter, the energy spread of the particle emission and the brightness for different sources for electron or ion emission. The data in brackets refer to helium ions with \unit[1]{mbar} ionization gas pressure introduced into the vacuum chamber.}
	\label{tab:comparison}
\end{table*}

The fourth emitter type in Tab. \ref{tab:comparison} is the SAT which was first realized by Fu et al. \cite{Fu2001} and improved by Kuo et al. \cite{Kuo2006a,Kuo2008}. A sketch of this tip is shown in Fig. \ref{fig:tip}. It consists of an etched single crystal (111) tungsten wire where a monolayer of iridium is deposited \cite{Kuo2006a}. After installation into ultrahigh vacuum, it is resistively heated several times to a temperature between \unit[$\sim$ 1000]{} and \unit[1300]{K} for \unit[3]{minutes}. Due to surface energy optimization, the tip forms a three-sided atom pyramid with a single iridium atom at the apex. As determined in \cite{Kuo2008}, the nano-pyramid can be regenerated in vacuum over 50 times by annealing and the long time stability of these SAT sources is extraordinary high. Variations in the ion current of \unit[3]{\%} for helium and \unit[5]{\%} for hydrogen for about \unit[30]{minutes} have been measured and they do not show any degradation after a total operation time of \unit[80]{hours} \cite{Kuo2008}. As for the supertips, a self focusing effect limits the emission angle leading to a high beam brightness for helium ions of \unit[$2 \times 10^{15}$]{A/m$^2$ sr} for ionization gas with a partial pressure of \unit[1]{mbar} \cite{Kuo2008}. The opening angle $\alpha$ of a coherent emission can be assessed by the relation for the angular coherence constraint $d \cdot \alpha \ll \frac{\lambda_{dB}}{2}$ \cite{Nicklaus1989,Mollenstedt1956a} where $d$ is the source size and $\lambda_{dB}$ the de Broglie wavelength of the particle. A SAT field emitter yields the smallest source size possible. The emitting area is only a single atom, with a diameter of \mbox{$\sim$ \unit[0.3]{nm}} in case of an iridium covered SAT \cite{Kuo2008}. This results in a large $\alpha$ and in a broad coherently illuminated area on the biprism and the detector. It is therefore well suited for electron as well as for ion interferometry with short de Broglie wavelengths.

\begin{figure}[htbp]
\centerline{\scalebox{0.17}{\includegraphics{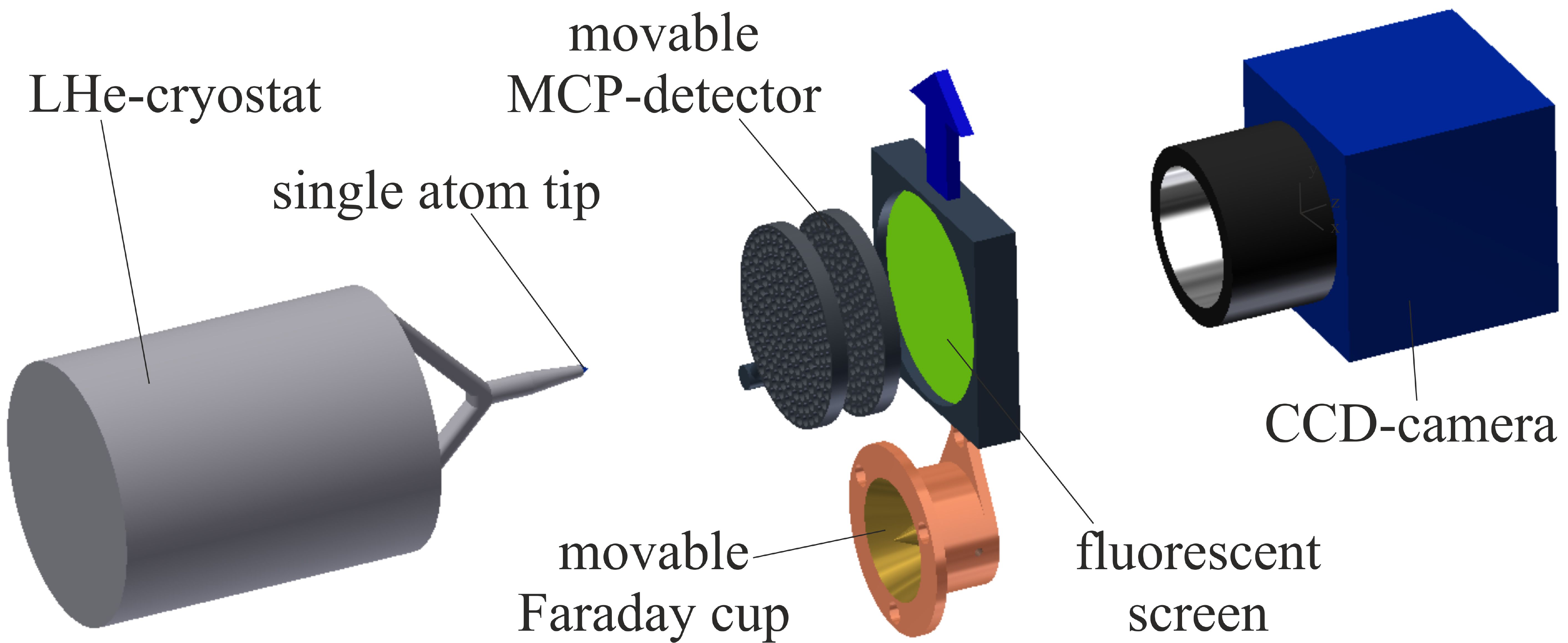}}}
\caption{(Color online) Sketch of the field electron or field ion microscope to characterize the SAT (not to scale). The electrons or ions emitted from a SAT source can be spatially resolved in a MCP detector in combination with a fluorescent screen and a CCD camera. For the counting of high signal rates the detector can be moved upwards to position a Faraday cup into the beam path.} 
\label{fig:testsetup}
\end{figure}

\begin{figure}[htbp]
\centerline{\scalebox{0.9}{\includegraphics{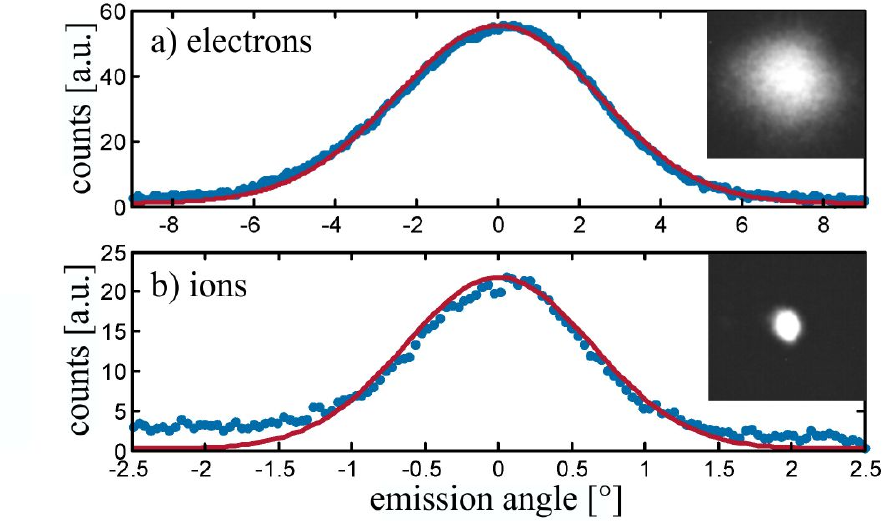}}}
\caption{(Color online) Emission pattern of an iridium covered W(111) SAT for a) electrons and b) helium ions. The images in the insets are to scale and show the spatial distribution of the emission recorded by the CCD camera. The data in the histograms is the signal integrated along the vertical direction (blue dots) overlapped by a Gaussian fit curve (red line).} 
\label{fig:electron_emission}
\end{figure}

\section{Experiment}
\label{experimental}

\subsection{Characterization of the SAT Source}

\begin{figure}[t]
\centerline{\scalebox{1.0}{\includegraphics{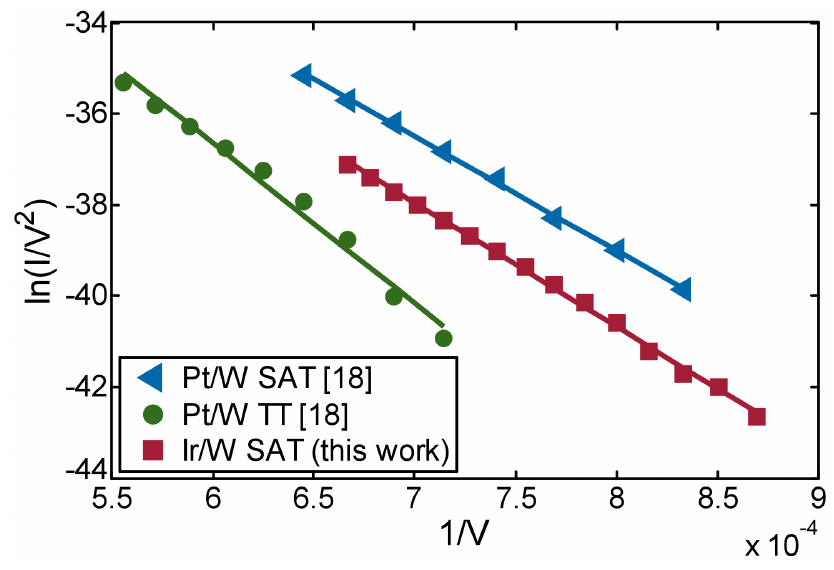}}}
\caption{(Color online) Fowler-Nordheim plot of the measured electron emission from the iridium covered W(111) SAT (red squares). The data is accompanied by a linear fit curve according to the Fowler-Nordheim equation  \cite{Fowler1928,Forbes1999} and compared to the emission from a platinum SAT (blue triangles) and a platinum trimer tip (green dots) taken from the literature \cite{Kuo2006a}.} 
\label{fig:FN_plot}
\end{figure}

The emission of an iridium covered W(111) SAT was characterized after annealing and forming in the field electron/ion microscope \cite{Tsong1990} illustrated in Fig. \ref{fig:testsetup} in the ultrahigh vacuum of \unit[$1\times10^{-9}$]{mbar}. The tip temperature during annealing was measured through a window from outside the vacuum chamber by a disappearing filament pyrometer. To observe the electron field emission, the tip was set on room temperature and on a voltage of \mbox{\unit[-1050]{V}}. The resulting spatial distribution of the emission pattern after a distance of \unit[42]{mm} behind the tip was measured in a MCP detector with a fluorescent screen. The images were recorded with a CCD camera outside the vacuum. A typical pattern is shown in the inset of Fig. \ref{fig:electron_emission} a) together with a histogram of the signal integrated along the vertical direction. The emission angle was measured to be 6.15 ($\pm \, 0.04$)$^\circ$ taking the full width at half maximum beam diameter. This is in good agreement with measurements for a platinum covered SAT from the literature with an electron emission angle of 5.6$^\circ$ \cite{Kuo2004}. Increasing the voltage soon saturates the MCP detector. Therefore it can be moved out of the beam path by a translation stage to be exchanged by a Faraday cup. It collects the emitted electrons and the charge is read out by a picoamperemeter. With this device a Fowler-Nordheim plot \cite{Fowler1928,Forbes1999}, shown in Fig. \ref{fig:FN_plot}, was recorded for field emission voltages between -1150 and \unit[-1500]{V}. It is compared to data from literature for a platinum single atom and a platinum trimer tip \cite{Kuo2006a}. Our SAT data was linearly fitted according to the Fowler-Nordheim equation where the slope $S$ of the fit in Fig. \ref{fig:FN_plot} can be calculated to be $S=b\,\Phi^{3/2} / \beta$ \cite{Fowler1928,Forbes1999}. With the work function $\Phi=$ \unit[5.03]{eV} for iridium and the constant $b=4\,(2m_e)^{1/2}/(3e\hbar)$ a field proportionality constant of \unit[$2.8\times 10^6$]{m$^{-1}$} was determined. As can be seen in Fig. \ref{fig:FN_plot}, our data agree well with the plots for the Pt/W(111) SAT from the literature which verifies the single atom emission of our Ir/W(111) SAT. \\
\\
We also measured the ion emission from the iridium covered W(111) SAT for different ionization gases. The tip was cooled down to \unit[38]{K} by a liquid helium cryostat. Helium ionization gas was streamed in the chamber until the pressure raised up to \mbox{\unit[4 $\times 10^{-7}$]{mbar}}. At an applied positive tip voltage of \unit[3.39]{kV} the helium field ionization pattern was measured with the MCP detector. The image is shown in the inset of Fig. \ref{fig:electron_emission} b) together with a histogram of the signal integrated along the vertical direction. A full width at half maximum emission angle of 1.48 ($\pm \, 0.06$)$^\circ$ could be determined. We recorded similar images also for hydrogen and neon ionization gas. The beam divergence is significantly smaller than for electrons. The measured spread is in perfect agreement with data from the literature for an iridium covered W(111) SAT where an emission angle of 1.5$^\circ$ was measured for helium at a tip temperature of \unit[20]{K} and a voltage between 7 and \unit[8]{kV} \cite{Kuo2008}. It needs to be noted that the ion emission voltage is strongly dependent on the diameter of the tungsten tip below the iridium layer (see Fig. \ref{fig:tip}) and can vary by several kV for different tips as well as for the same tip after annealing to regenerate the single atom emission. \\
In the cited literature \cite{Kuo2006a,Kuo2008} the pyramidal atomic structure was verified by field evaporation of the upmost iridium atom. As a result it could be observed that the helium gas gets ionized by the three iridium atoms forming the second layer of the nano-pyramid. Another field evaporation revealed the third layer with 10 iridium atoms visible in the field ion microscope. In the interpretation of our data we rely on these work and reason from the equivalence of our determined ion emission angle with the value in \cite{Kuo2006a,Kuo2008} that we achieved single atom tip emission where only the topmost atom is the origin of the particle beam for ions and electrons.

\subsection{A New Preparation Method for the Biprism Beam Splitter}
\label{biprism}

The electrostatic biprism is the second core element in the interferometer. It acts as a coherent beam splitter for the charged matter wave and requires to be electrically conducting. In electron and ion interferometers \cite{Hasselbach1998a,Hasselbach1996,Maier1997,Hasselbach2010a,Sonnentag2007a,Mollenstedt1956a,Nicklaus1993a,Hasselbach1988a} it usually has a surface roughness in the nanometer range and a diameter below \unit[1]{\textmu m}. The former common technique to manufacture such thin wires was to manually draw a glass fiber out of a silica rod above a hydrogen-oxygen flame \cite{Maier1997,Mollenstedt1956a}. The technique needed a lot of experience and could produce wires with typical diameters between 200 and \unit[600]{nm} \cite{Maier1997}. The drawback was the diameter variation of up to several hundred nanometers between different preparations. Therefore, no exact information about the diameter was available, leading to an uncertainty that enters in the analysis of the interference data. A prior measurement of the diameter in a scanning electron microscope (SEM) was not possible, since the biprism got contaminated with dust during the pumping procedures in the apparatus. If dust deposits get charged by the electron beam, they deform the interference pattern. 

\begin{figure}[htbp]
\centerline{\scalebox{1.0}{\includegraphics{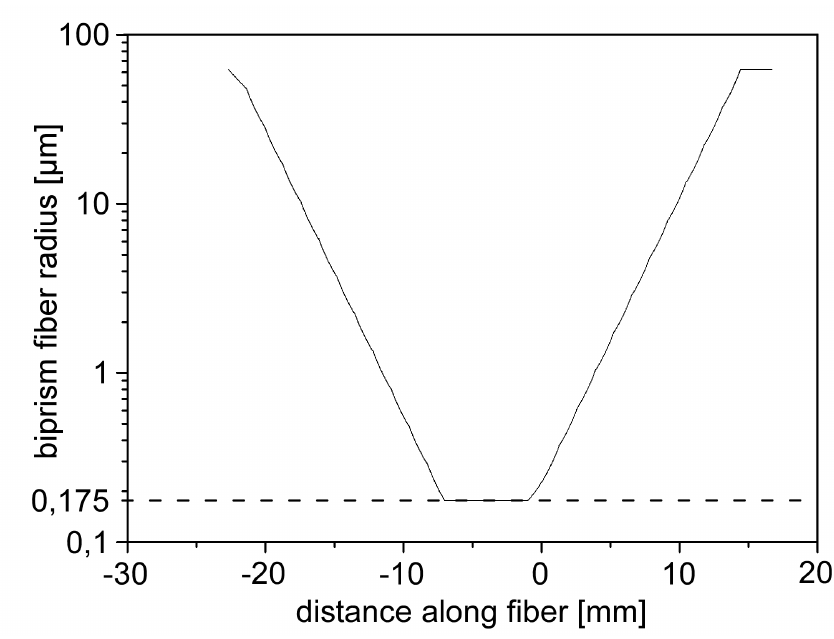}}}
\caption{Nominal fiber radius-profile of the uncoated biprism wire for the interferometer.} 
\label{fig:figure_neu}
\end{figure}

\begin{figure}[t]
\centerline{\scalebox{0.61}{\includegraphics{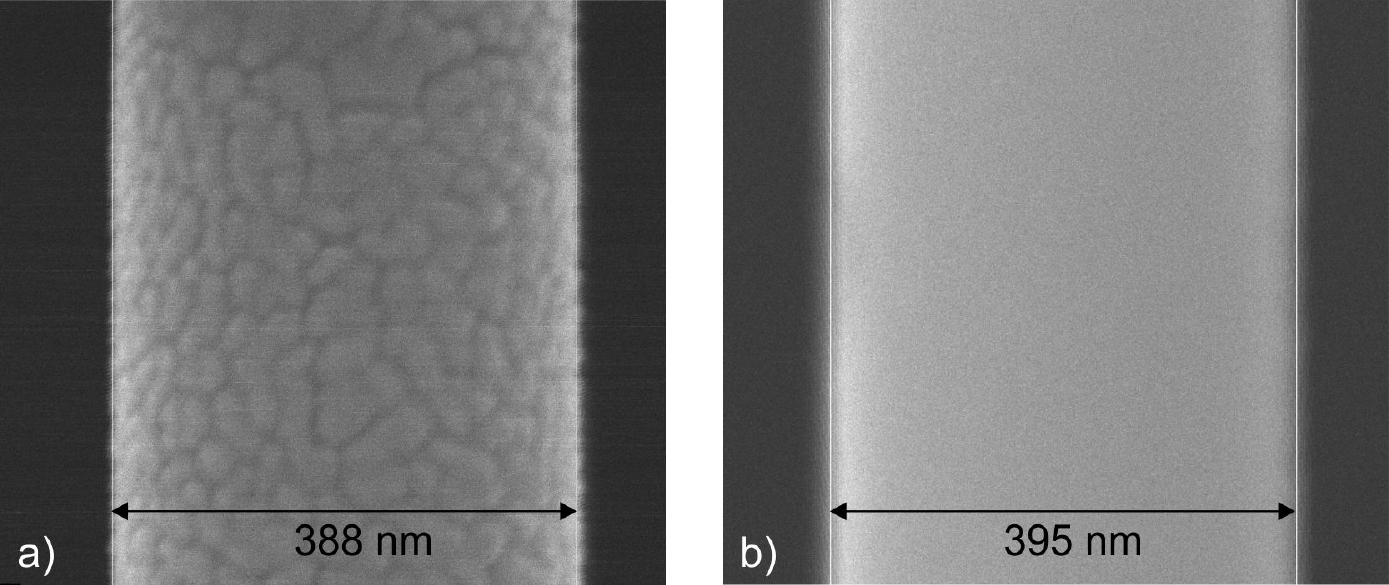}}}
\caption{SEM images of two biprism fibers based on silica nanowires drawn with the technique described in the text \cite{Warken2007,Warken2007b}. The fiber in a) was coated with gold and in b) with gold-palladium (60:40).} 
\label{fig:fiber}
\end{figure}

For that reason we developed a new biprism preparation procedure where a certain diameter can be chosen and reliably reproduced without further SEM analysis of each newly created fiber.
The method relies on a custom-built computer-controlled fiber-pulling rig \cite{Warken2008,Warken2007,Warken2007b}. For the heat-and-pull process a standard optical fiber was attached on two linear translation stages and positioned above a hydrogen-oxygen flame with a stoichiometric gas mixture. The softened fiber was stretched and at the same time translated with respect to the flame along its axis. With that technique tapered optical fibers with pre-defined shape can be created \cite{Stiebeiner2010}. For the biprism beam splitter fabricated here this radius profile consists of a nanowire section with a homogeneous nominal diameter of \unit[350]{nm} and a length of \unit[6]{mm} that is linked to the unprocessed fiber ends by two tapered fiber sections. The corresponding nominal radius profile is presented in Fig. \ref{fig:figure_neu}. Using SEM measurements, it was shown that the actually realized fiber-profiles deviate by less than \mbox{\unit[$\pm \, 10$]{\%}} from the intended design \cite{Stiebeiner2010}. With this technique it is possible to produce fibers with a diameter down to about \unit[100]{nm} \cite{Stiebeiner2010}. In order to test if this procedure can be applied for the creation of an interferometer biprism, two fibers were pulled as described and fixed on an U-shaped titanium mount with ultra-high vacuum compatible glue. Thin layers of gold and a gold-palladium (60:40) alloy were subsequently deposited on the fibers with equal sputtering times. To prevent effects related to contact potentials, the holders and the grounded electrodes were sputtered as well \cite{Maier1997}. To ensure a constant coverage on all sides of the biprism, the probes were turned during the sputtering process. Fig. \ref{fig:fiber} shows SEM images of the resulting biprism wires. The diameters including the coatings inferred from these images are \unit[388]{nm} for gold and \unit[395]{nm} for gold-palladium. Based on the nominal diameter of the silica nanowires of \unit[350]{nm} the layer thicknesses was estimated to be \unit[19]{nm} and \unit[23]{nm}, respectively. The uncertainty in these values are not identified. A low surface roughness is an important requirement for the biprism wire to be used as a beam splitter in interferometry, since irregularities lower the fringe contrast. As inferred from the image the gold-palladium alloy surface is clearly more smooth than the gold surface.

\begin{figure*}[htbp]
\centerline{\scalebox{0.5}{\includegraphics{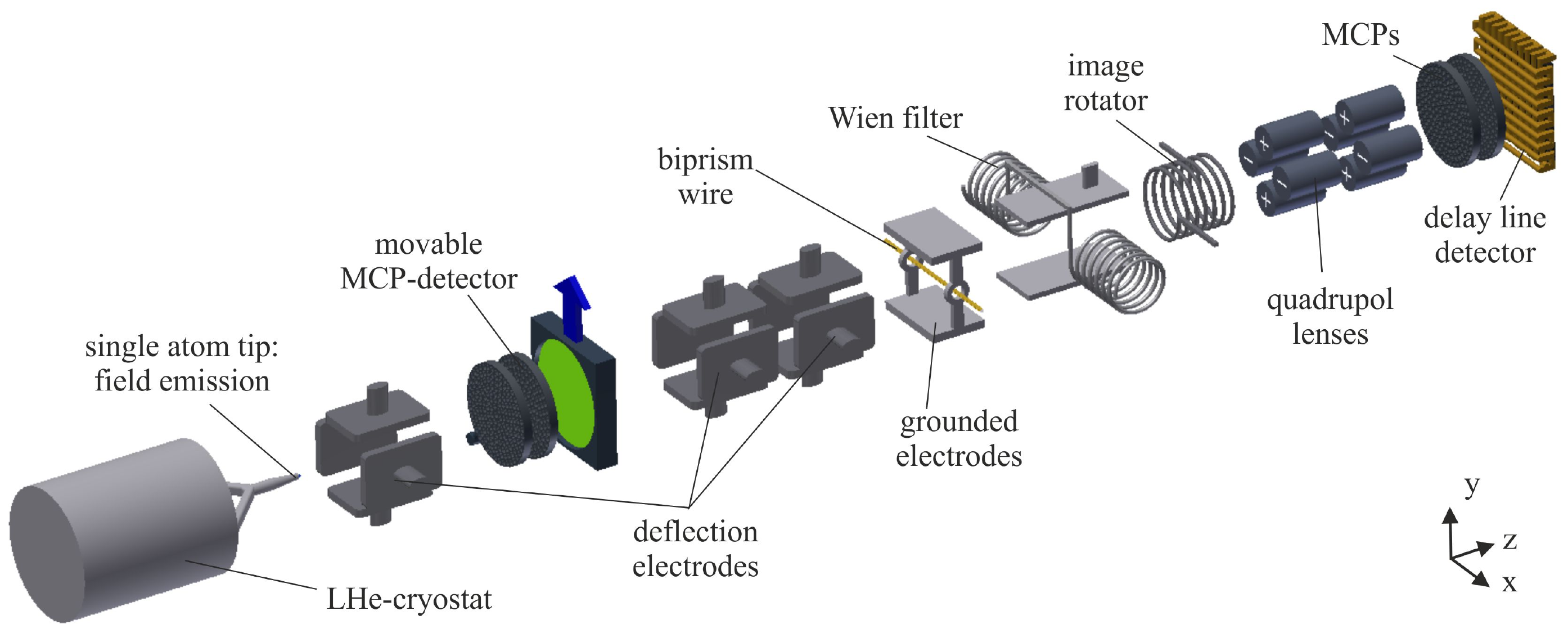}}}
\caption{(Color online) The experimental setup of the interferometer. It was first realized by Hasselbach et al. \cite{Hasselbach1998a,Hasselbach1996,Maier1997,Hasselbach2010a} and modified for this study. The changes concern mainly the beam source, the biprism and the detector.} 
\label{fig:setup}
\end{figure*}

\subsection{An electron biprism interferometer with a single atom tip source}

To demonstrate electron interference with the SAT source, the setup illustrated in Fig. \ref{fig:setup} was used. The interferometer was constructed by Hasselbach et al. \cite{Hasselbach1998a,Hasselbach1996,Maier1997,Hasselbach2010a} and modified to be applied for this study. Relevant changes concern the beam source, the biprism and the detector. The supertip beam source was replaced by a SAT. The biprism beam splitter was exchanged by a fiber drawn according to the method described above. And the MCP detector with a phosphor screen and a CCD camera was substituted by a delay line detector. All parts are in a vacuum chamber with a pressure of \unit[$\sim 4\times 10^{-9}$]{mbar}. In the compact interferometer architecture \cite{Hasselbach1988a} all components are mounted on two rigid ceramic bars and the ion beam is aligned by three deflection electrodes. It is shielded against external electric and magnetic fields by a copper and a mu-metal tube. The electron source is an Ir/W(111) SAT \cite{Kuo2008} that was prepared according to the procedure described above. It can be cooled by a liquid helium cryostat for future applications with ions. After the first deflection electrode for beam aligning the particles get detected by a movable MCP detector. This part together with a phosphor screen forms a field electron or field ion microscope \cite{Tsong1990} which is important to observe the formation of the SAT in the vacuum after annealing. With the help of a $\unit[45]{^\circ}$ mirror, the screen can be seen through a side window in the chamber. The MCP detector can be moved out of the optical axis to align the beam onto the biprism by two further deflection electrodes. The biprism separates the beam coherently and overlaps the two partial waves. It is positioned in the middle between two grounded electrodes that are \unit[4]{mm} apart from each other. The Wien filter \cite{Nicklaus1993a} consists of two magnetic coils in combination with two electrodes. It is able to correct the longitudinal wave package shift caused by the aligning deflectors if the beam path does not perfectly match the optical axis. After the partial matter waves combine, they form an interference pattern which is oriented along the $x$-axis of the quadrupole lenses by the image rotator. It is a coil inducing a magnetic field parallel to the direction of the beam propagation. The quadrupole lenses magnify the image in the $y$-direction by a factor up to 10$^5$ \cite{Maier1997} to fit the detectors resolution of about \unit[100]{\textmu m}. At the end the particles get detected by a combination of a double-stage MCP and a delay line anode \cite{Jagutzki2002}. It allows for a spatial and high temporal resolution with a dead time between two pulses of \unit[310]{ns} and a total time accuracy below \unit[1]{ns}.\\

\begin{figure*}[htbp]
\centerline{\scalebox{1.0}{\includegraphics{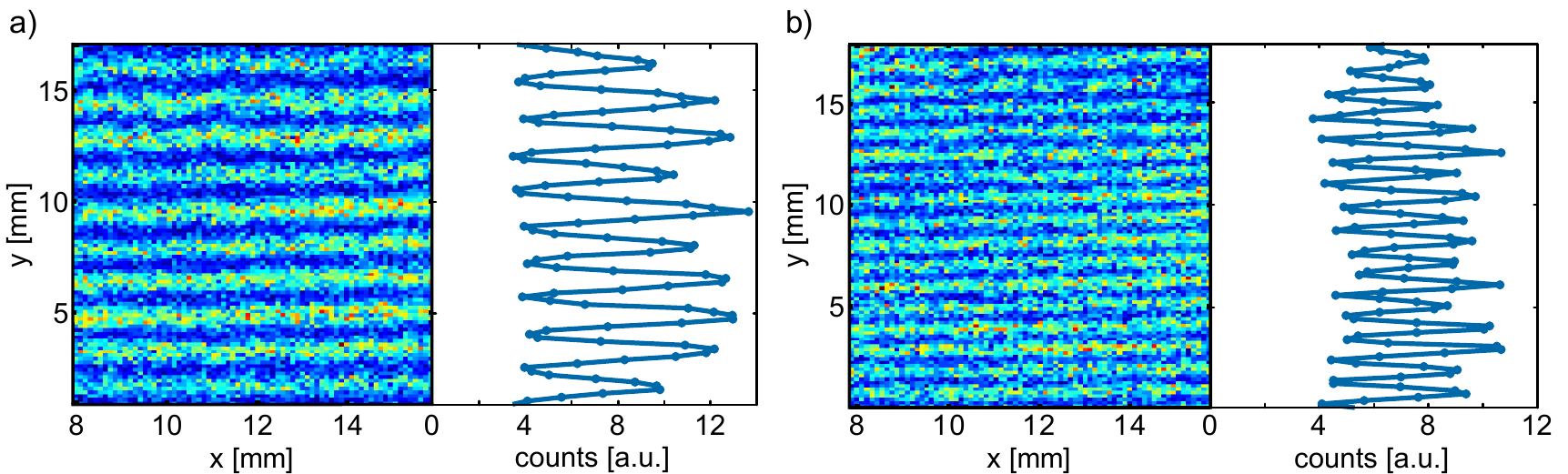}}}
\caption{(Color online) Electron interference pattern on the detector using a SAT source. The count rate was $\sim \unit[3]{kHz}$. In a) the biprism voltage was set to \unit[0.5]{V} and in b) to \unit[0.7]{V} leading to different diffraction angles and smaller fringe distances. On the right side of each image the integrated signal along $x$ is plotted (blue dots). .} 
\label{fig:interferences}
\end{figure*}

\begin{figure}[htbp]
\centerline{\scalebox{0.9}{\includegraphics{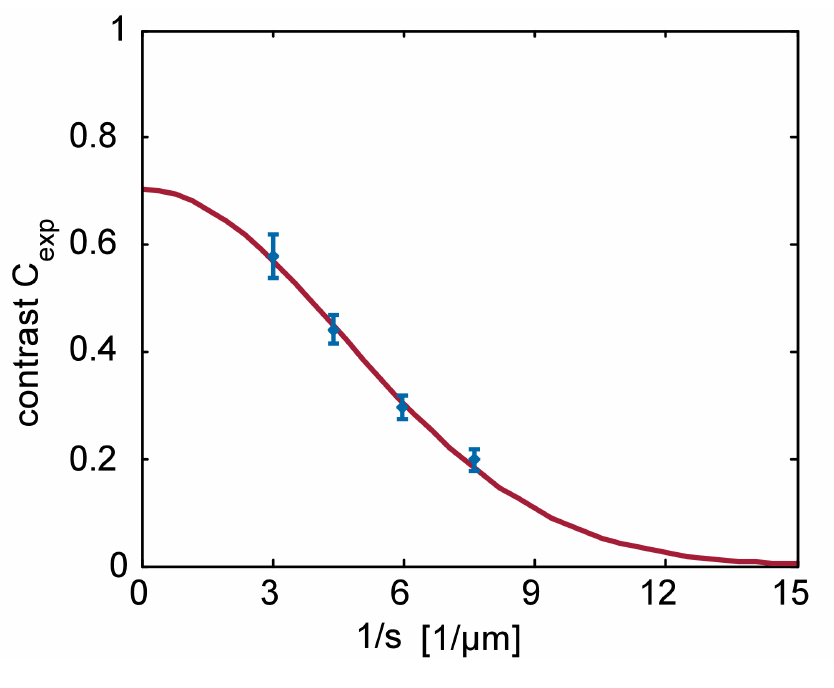}}}
\caption{(Color online) Diagram of the determined interference contrast $C_{exp}$ versus the inverse fringe distance $s$ from interference patterns with four different biprism potentials: 0.5, 0.7, 0.9 and \unit[1.1]{V} (blue dots). The data is fitted by Eq. \ref{equ:contrast} (red line).} 
\label{fig:contrast}
\end{figure}

\begin{figure}[htbp]
\centerline{\scalebox{1.0}{\includegraphics{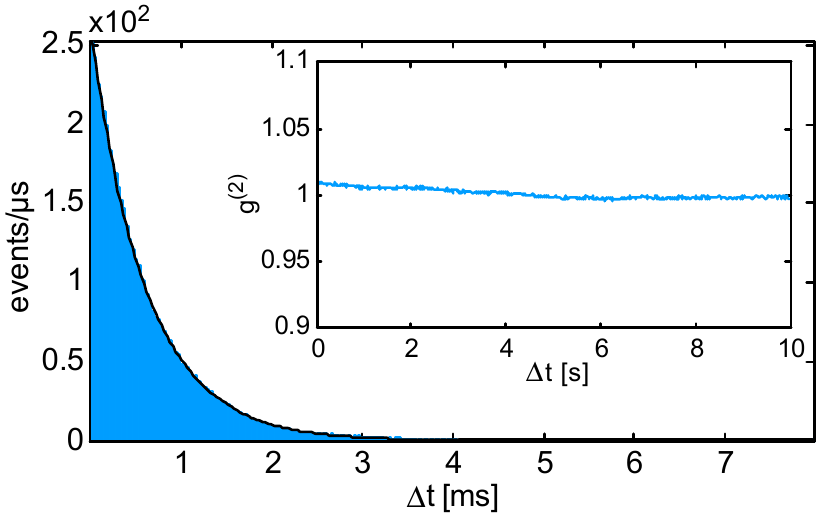}}}
\caption{(Color online) Histogram of the time distances between subsequent electron events in Fig. \ref{fig:interferences} b) together with a calculation assuming Poisson-distributed emission (black line). Inset: The second order correlation function  $g^{(2)}$ for detected electron pairs which have a time distance from 0 to \unit[10]{s} for the whole data of Fig. \ref{fig:interferences} b).} 
\label{fig:time}
\end{figure}

Two resulting interference pattern for electrons with an acceleration voltage of \unit[1.55]{kV} are demonstrated in Fig. \ref{fig:interferences} using the gold-palladium covered biprism on a potential of \unit[0.5]{V} in a) and \unit[0.7]{V} in b). The contrast is determined to be \unit[51.3 ($\pm \, 4.1$)]{\%} and \unit[35.8 ($\pm \, 3.1$)]{\%}, respectively. We took similar images for other biprism potentials and in a separate series of measurement for the gold covered biprism fiber. In each image the interference contrast $C_{exp}$ and the fringe distance $s$ was determined. The interference contrast depends on the lateral and longitudinal coherence of the source. The first one is determined by the source width, the second by the wave number distribution and therefore by the energy spread of the particle emission \cite{Lenz1984}. Due to the small energy spread of the SAT (see Tab. \ref{tab:comparison}), the latter contribution can be neglected. Therefore, the theoretical interference contrast depends only on the source width and can be obtained by the expression \cite{Lenz1984}:

\begin{equation}
C_{th} = 1/\gamma \, C_{exp} = exp\left(-2\left(\frac{\pi\, M\, \eta\, \epsilon}{p \,s}\right)^2\right) \label{equ:contrast}
\end{equation}

whereby the factor $\gamma$ considers a reduction of the theoretical contrast by effects such as e.g. deflected incoherent background electrons, mechanical vibrations, temperature drifts or impurities on the biprism wire. The parameter $\epsilon$ is the width of the source which is in \cite{Lenz1984} assumed to be Gaussian distributed, $p$ is the distance between the source and the biprism and $\eta$ the one between the biprism and the entrance of the quadrupole lens. We determined the quadrupol magnification $M\sim 4460$ in a computer beam path simulation with the program Simion. In Fig. \ref{fig:contrast} the measured values for the contrast $C_{exp}$ are compared with the fringe distance $s$ after magnification for different biprism potentials. By a fit of Eq. \ref{equ:contrast} a reduction factor of $\gamma=0.72$ ($\pm \, 0.05$) and a source size of \mbox{\unit[$\epsilon = 37$ ($\pm \, 2$)]{nm}} was determined for the gold covered biprism fiber. In a further series of measurement with the gold-palladium covered fiber we extracted a value of \mbox{\unit[$\epsilon = 44$ ($\pm \, 5$)]{nm}}. Therefore the emission spot determined by the application of this theory is significantly larger than the size of an iridium atom (\unit[$\sim 0.3$]{nm}). The reason for this deviation might be related to an oscillation of the biprism fiber. The effect of such a motion is qualitatively equivalent to a motion of the source. It leads to a temporary varying phase shift of the single particle waves and to the same decrease in contrast as for a broader emission center. An evidence for the vibration amplitude was given by an experiment in a different field of research, but with a comparable fiber. By the evanescent coupling between a bottle microresonator and the subwavelength-diameter waist of a tapered optical fiber, the root mean square fluctuation of the fiber was estimated to be \unit[$\pm \, 9$]{nm} in the frequency range up to \unit[200]{Hz} \cite{Junge2011}. Such a biprism vibration would lead to an interference contrast related to a source size of $\sim$ \unit[18]{nm}, even for SAT emission. The thermal amplitude of the string vibration of the fiber in \cite{Junge2011} was calculated to be \unit[$\pm \, 5$]{nm}. This could be a limit of source size resolution by the contrast in a biprism interferometer. However, the low frequency contribution of this vibrational dephasing can in principle be removed by a correlation analysis of the spatial and temporal signal in the detector \cite{Rembold2013}. \\
\\
We also recorded the emission characteristics of the SAT in the time domain. The histogram in Fig. \ref{fig:time} shows the distribution of the time distances $\Delta t$ between consecutive electrons from the data in Fig. \ref{fig:interferences} b). It is superposed with a calculation (black curve) assuming a Poisson-distributed SAT emission that follows the relation $E=(\nu e^{-\nu \Delta t}) / \tau$ with the count rate $\nu$ and the binning width of our data representation in the histogram $\tau$. The calculation agrees perfectly with the measurement. The inset of Fig. \ref{fig:time} exhibits the second order correlation function $g^{(2)}$ of all particle pairs which have a time difference $\Delta t$ between 0 and \unit[10]{s} for the whole measurement of the data in Fig. \ref{fig:interferences} b). It was extracted and calculated according to a scheme described in \cite{Rembold2013}. The determined deviation to $g^{(2)}=1$ of less than \unit[1]{\%} verifies a Poisson-distribution of the SAT emission between all detected electron pairs with a time distance from 0 to \unit[10]{s}.

\section{Conclusion}

We demonstrated the first biprism electron interferometer with a single atom tip field emitter in a modified setup based on \cite{Hasselbach1998a,Hasselbach1996,Maier1997,Hasselbach2010a} and analyzed the interference pattern for different superposition angles. The new source was also characterized in a separate field electron and field ion microscope. It was demonstrated that the emission spot size is only a single iridium atom resulting in a beam of high spatial coherence that is emitted into an angle of only \mbox{6.15 ($\pm \, 0.04$)$^\circ$} for electrons and \mbox{1.48 ($\pm \, 0.06$)$^\circ$} for helium ions due to a self focusing effect related to the tip geometry. The field emission of the Ir/W(111) SAT was also characterized in the time domain by the second order correlation function. It verified that the emission is perfectly Poisson-distributed with a deviation of less than \unit[1]{\%} for time distances from 0 to \unit[10]{s} between the detected electrons. We provided a comparison of important beam emission features with other electron and ion sources indicating the superior properties of single atom tips as a novel source for biprism electron and ion interferometers.\\
Furthermore, we presented a method for biprism wire fabrication to minimize the fiber-diameter uncertainty. The biprism is drawn from a standard optical fiber in a highly controlled and reproducible heat-and-pull process. The resulting silica nanowire was then coated with a thin gold and gold-palladium layer, yielding a diameter of about \unit[400]{nm} and a length of \unit[6]{mm}.\\
Therefore, our setup has the improvement of a well defined source and biprism fiber diameter in contrast to former biprism interferometers \cite{Hasselbach1998a,Hasselbach1996,Maier1997,Hasselbach2010a,Sonnentag2007a,Mollenstedt1956a,Nicklaus1993a,Hasselbach1988a}.  Combined with the extremely small source size of \mbox{$\sim$ \unit[0.3]{nm}}, this allows for a more accurate analysis of the interference data.
Although the field emission pattern and a Fowler-Nordheim plot indicate single atom tip emission, the analysis of the interference contrasts result in a source size of \mbox{\unit[37 ($\pm \, 2$)]{nm}}. We assume that this outcome is due to string vibrations of the biprism fiber that dephase the matter waves and result in a lower contrast. The setup has applications for future ion interference experiments in Aharonov-Bohm physics such as the first direct measurement of the electric Aharonov-Bohm effect \cite{Schutz2013}.\\
\\

\section{Acknowledgements}

This work was supported by the Deutsche Forschungsgemeinschaft (DFG, German Research Foundation) through the Emmy Noether program STI 615/1-1. A.R. acknowledges support from the Evangelisches Studienwerk e.V. Villigst. The authors thank H. Prochel and F. Hasselbach for helpful discussions.

\end{document}